\documentstyle[pra,aps,amssymb,tighten,epsfig,multicol]{revtex}
\newcommand{\twoket}[2]{\ket{{#1}}\!\!\ket{{#2}}}
\newcommand{\twobra}[2]{\bra{{#1}}\!\!\bra{{#2}}}
\newcommand{\ket}[1]{\left| {#1} \right\rangle}
\newcommand{\bra}[1]{\left\langle {#1} \right|}
\newcommand{\op}[1]{\hat{#1}^{}{}}
\begin{document}
\title{Quantum copying can increase the practically available information}
\author{P. Deuar\cite{PDemail} and W. J. Munro}
\address{Centre for Laser Science,Department of Physics, University of
Queensland, QLD 4072, Brisbane, Australia}
\date{\today}
\maketitle
\begin{abstract}
  While it is known that copying a quantum system does not increase the
amount of information obtainable about the originals, it may increase the
amount available in practice, when one is restricted to imperfect measurements.
We present a detection scheme which using imperfect detectors, and
possibly noisy quantum copying machines (that entangle the copies), allows one to extract more
information from an incoming signal, than with the imperfect detectors
alone. The case of single-photon detection with noisy, inefficient
detectors and copiers (single controlled-NOT gates in this case)  is investigated in detail. The improvement
in distinguishability between a photon and vacuum is found to occur for
a wide range of parameters, and to be quite robust to random
noise. The properties that a quantum
copying device must have to be useful in this scheme are investigated.
\end{abstract}
\pacs{}

\begin{multicols}{2}
\section{Introduction}
\label{INTRO}

  It is well known that making copies of a quantum system (e.g. with a quantum
copier) does not increase the amount of information present about the
original. To put it another way, spreading information about the
original system onto several systems does not increase the amount of
information that one can obtain about the original (in fact, this usually
decreases it due to noise). However, in discussions on this matter it
is usually tacitly assumed that one has access to optimal measuring
devices. 

  In practical situations, however, this is never the case. One is
always restricted to imperfect measurements, due to inefficient
detectors, and various sources of random noise.  
  Although, in theory, Quantum mechanics allows one to perfectly
distinguish between orthogonal states by making appropriate
measurements, in practice distinguishing perfectly every time is
impossible. 
 Of course, in many situations these imperfections of measurement
 are insignificant, but in this article we consider those 
cases where such inefficiencies are relevant. 
  
Let us investigate what can be done in principle if one is
restricted to using inefficient and noisy detectors. In many practical
situations what one is interested in is to determine in which one of
several possible orthogonal states a system is residing. For example,
this is what one does to extract transmitted information from a
signal. 

  The basic idea explored in this article can be expressed as follows:
If we can get a second chance to use the detectors at our disposal on
the same state, we might do better at distinguishing it from among the
range of possibilities. We will investigate what happens when one
makes copies of the original state. If the available detectors are
fairly poor, then one may hope that making even imperfect copies may
still give improvements if one can then make independent measurements
on each of the copies.

     Copying machines in general use two approaches. One of the extreme cases
is a classical copying machine, where measurements (destructive or
non-destructive) are made on the original state, the results of which
are then fed as parameters into some state preparation scheme that
attempts to construct a copy of the original. This approach obviously
allows one to generate an arbitrary amount of copies, possibly all identical to
each other.  The opposite extreme is a fully quantum copying machine
that by some process that is unseen by external observers (a ``black box''), creates a
fixed number of copies, usually destroying the original in the
process. Naturally in a realistic situation, noise will additionally
degrade the quality of the copies, and copiers that utilize both of the 
processes above are obviously also possible.

Since one's detection resources are restricted to imperfect detectors that discard some
information about the state, then it becomes
immediately obvious that classical copying gains you nothing. Any
information about the original state that you can extract from the
copies can  be extracted just as well from the measurement results used to produce
the copies -- and these are made with those imperfect detectors.
Quantum copying, however, \emph{is} able to give improvements, even
when degraded by noise and inefficiencies, as will
be seen below.

For simplicity, and because the aim is above all to demonstrate the
principle at work here, we will consider situations where one wishes
to distinguish between two orthogonal possibilities for the input
state. Some examples of this would be single-photon detection,
distinguishing spins of spin-half particles, single-photon
polarization, or distinguishing between some number of photons and no
photons. 

This paper sets out in more detail, and expands on a previous short
article dealing with this topic  by the same authors\cite{summ}. 
Sec.~\ref{THESCHEME} puts forward the general detection scheme that,
utilizing entangling quantum copiers and inefficient detectors, allows one (if
the copiers are good enough) to
achieve surer detection than with the detectors alone. An example is
given with a very simplified case of single-photon detection.
Sec.~\ref{THEMODEL} Develops a more realistic schematic model of
single-photon detection, using a single controlled-NOT gate as the
copier. 

Subsequently, in Sec.~\ref{PERFORMANCE} we consider the noiseless
case and analyze its performance with respect to the standard
one-detector setup. We first consider the situation where one  uses
the measurement results to make a decision about what the original
state was -- the probability of being correct is compared between
detection schemes. Secondly, we compare the total information about
the original state that is  in
principle extractable from the measurement outcomes.
Sec.~\ref{NOISE} Looks at how robust the copier-enhanced detection
scheme is to random noise in the copiers and detectors.
Finally, in Sec.~\ref{PROP} the properties that a quantum
copying device must have to be useful are found.

\section{A Detection Scheme with Quantum Copiers}
\label{THESCHEME}
\begin{figure}
\center{\epsfig{figure=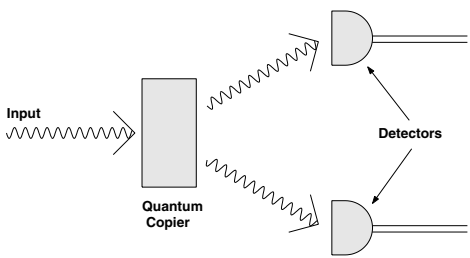,width=70mm}}
\caption{Basic detection scheme using imperfect detectors, and a quantum copier}
\label{SCHEMEfig}
\end{figure}

Consider the case where one of a set of possible input states are
to be distinguished by a measurement scheme, using (some number of
identical) imperfect
detectors. That is, whether the input states are mutually orthogonal,
or not, the detectors at one's disposal do not always distinguish between the
inputs with certainty. One also has some (identical) quantum copiers that can act
on the possible input states. For a first look at the situation, let us suppose that the
possible input
states are mutually orthogonal, and that one has somehow acquired
perfect quantum copiers for this set of states. Assume the copiers
destroy the original, and produce two copies for simplicity. Then, an obvious way
to take advantage of the copiers is to send the originals through a
quantum copier, before trying to detect both copies separately. (As in
Fig.~\ref{SCHEMEfig}). This basically gives one a second chance to
distinguish the input state, if the detection at the first copy fails.

 In practice, one can never be certain whether the result given
by a detector is due to noise, or the input state, but in this case,
having two tries at detection allows one to better estimate whether
the result was trustworthy - once again on average increasing one's
knowledge of the original. 

To be slightly more concrete, consider a very simplified model of
photodetection using this measurement scheme. (A more realistic model
is developed in Sec.~\ref{THEMODEL}).  Suppose one has perfect
copiers, and noiseless photodetectors of efficiency $\eta$. That is, the probability
of a a count on the detector is $\eta$ if a photon is incident, and
$0$ otherwise. 

With the copier set up as in Fig.~\ref{SCHEMEfig}, if any of the
detectors register a count, one can with certainty conclude that a
photon was incident. So, if a photon \emph{is} incident, the
probability of finding it is 
\begin{equation}
P^{(1)}_{\text{count}|\text{photon}} = 
\eta+(1-\eta)\eta,
\end{equation}
 as opposed
to just 
$\eta$ with no copier, 
  because one gets a
``second chance'' at detection. On the other hand, if no count is
registered, then the probability that 
no photon was incident is
\begin{equation}\label{Boo}
P^{(1)}_{\text{nophoton}|\text{nocount}} =
\frac{1-p}{1-\eta p(2-\eta)},
\end{equation}
where $p$ is the probability
that a photon is incident on average, irrespective of the measurement
result. The expression of Eq.\ (\ref{Boo}) is always greater than
$(1-p)/(1-\eta p)$, which is the probability if no copier is
used. This increase reflects the added confidence that comes from
both detectors failing to register the photon. 

With more copiers, one can do better. Instead of placing
photodetectors at the outputs of the first copier, place copiers instead
, and detect photons only when they have come out of the second lot of
copiers. One can continue putting in more copiers in a similar
fashion.  If we let the number of copiers that photons must
pass through before being detected be $N$,  ($N=1$ in the case
considered previously) then one finds that for this simplified scheme
\begin{mathletters}\label{Psimple}
\begin{eqnarray}
P^{(N)}_{\text{count}|\text{photon}} &=& 1 - (1-\eta)^{(2^N)},\\
P^{(N)}x_{\text{nophoton}|\text{nocount}} &=& \frac{1-p}{1-p+p(1-\eta)^{(2^N)}}.
\end{eqnarray}\end{mathletters}
  So as $N$ increases, the probability of detecting a photon that is present
(given that it is present) approaches one. Also, the probability that no photon
was present if it was not detected also approaches one.

Note that using quantum copiers, and not classical
ones is vital. A classical copier would have to rely on the same
imperfect 
photodetectors, and would actually \emph{reduce} the
detection efficiency, since to detect a photon at one of the two copy
detectors, one must have been first detected at the copier. This gives
$P^{(1)}_{\text{count}|\text{photon}} = \eta^2(2-\eta)$, which is
always less than or equal to $\eta$
($P^{(0)}_{\text{count}|\text{photon}} = \eta$ is achieved without any
copiers at all).

\section{A Model of Improved Single-Photon Detection}
\label{THEMODEL}
Detection with the help of perfect quantum copiers, as briefly
discussed in the previous section,  is all very well,
but what happens when the equipment used is noisy, and not 100\%
efficient?  
Consider the following, more realistic, model of photodetection, using
the scheme outlined in Sec.~\ref{THESCHEME}.

The possible states that are to be distinguished are the vacuum
$\ket{0}$ and single photon $\ket{1}$ states. The \emph{a priori} probability
that the input state is a photon is $p$. 

A generalized measurement on some state ${\hat \rho}$ can be modeled by
a
positive operator-valued measure (POVM) $\op\{{A}_i\}$ \cite{Kraus:83,CavesD:94}
described by a set of $n$ positive operators $\op{A}_i$,
 such that $\sum_{i=1}^{n} \op{A}_i = \op{I}$, where $\op{I}$ is the
identity matrix in the Hilbert space of $\op{\rho}$ (and of the $\op{A}_i$).  The probability of
obtaining the $i$th result, by measuring on a state $\op{\rho}$ is then
\begin{equation}\label{Pi}
P_i = \mbox{Tr}\left[{\op{\rho}\op{A}_i}\right].
\end{equation}

Now suppose the photodetectors at one's disposal are noisy and have quantum
efficiency $\eta$. The effect of these can be modeled by the POVM
\begin{mathletters}\label{DETpovm}
\begin{eqnarray}
  \op{A}_+ =& \eta\ket{1}\bra{1} + \eta\xi\ket{0}\bra{0},& \\
  \op{A}_- =& (1-\eta)\ket{1}\bra{1} +(1-\eta\xi)\ket{0}\bra{0},&
\end{eqnarray}\end{mathletters}
where the operator $\op{A}_+$ represents a count, and the operator
$\op{A}_-$ the lack of one.
The parameter $\xi\in[0,1)$  controls the amount of noise. That is, $\xi\eta$ is the
probability that the photodetector registers a spurious (``dark'') count when no photon is
incident.

Model the quantum copier as one that has a probability $\varepsilon$ of
working correctly and producing perfect copies. Otherwise, the parameter
$\mu\in[-1,1]$ determines (in a somewhat arbitrary way) what is produced. This
can be written
\begin{mathletters}\label{COPYtransf}\begin{eqnarray}
\op{\rho}_1 = \twoket{1}{d}\twobra{1}{d} &\to
\varepsilon\twoket{1}{1}\twobra{1}{1} +(1-\varepsilon)\op{\rho}_N = \op{\rho}^1_1,\\ 
\op{\rho}_0 = \twoket{0}{d}\twobra{0}{d} &\to \varepsilon\twoket{0}{0}\twobra{0}{0} +(1-\varepsilon)\op{\rho}_N = \op{\rho}^1_0.
\end{eqnarray}\end{mathletters}
$\ket{d}$ is a dummy state, that is fed into the copier, and becomes
the second copy. It is included here to preserve unitarity in the
perfect copying case $\varepsilon=1$. The state produced upon failure of the
copier, $\op{\rho}_N$ is independent of the original, and is given
by
\begin{equation}\label{NOISEstate}
\op{\rho}_N = (1-|\mu|)\frac{\op{I}}{4} + \
\left\{ \begin{array}{cll}
\mu&\twoket{1}{1}\twobra{1}{1} & \text{ if }\mu>0 \\
|\mu|&\twoket{0}{0}\twobra{0}{0} & \text{ if }\mu\le0
\end{array}\right. .
\end{equation}
Here, $\op{I}/4$ is the totally random mixed state.
 So, for $\mu=0$ a totally random noise
state is produced upon failure to copy, for $\mu=-1$ vacuum, for
$\mu=1$ photons in both copies, and for intermediate values of
$\mu$ a linear combination of the three cases mentioned.
The case briefly considered in the previous section had the parameters
$\varepsilon=1$, $\xi=0$.

This model (Eq.\ (\ref{COPYtransf})) of the copier is an extension (to allow for inefficiencies)
of the
Wootters-Zurek copier, which has been extensively studied
\cite{WZ:82,BuzekH:96}. In the ideal case ($\varepsilon = 1$), with the
dummy input  state in the vacuum ($\ket{d} = \ket{0}$),  the
transformation is: 
\begin{equation}\label{cnott}
\twoket{0}{0} \to \twoket{0}{0} \qquad 
\twoket{1}{0} \to \twoket{1}{1}.
\end{equation}
This transformation can be implemented by the simplest of all quantum logic circuits,
the single controlled-NOT gate. These have  recently begun to be  implemented
for some systems (although admittedly not for single-photon systems), and are the subject of intense ongoing research,
because of their application to quantum computing. This means that
similar schemes to the one considered here may become experimentally
realizable in the foreseeable future. 

Note that the transformation\ (\ref{cnott}) can be also considered an
``entangler'' rather than a copier. Consider its effect on the
photon-vacuum superposition state 
\begin{equation}\label{superpos}
\frac{1}{\sqrt{2}}(\ket{0} +\ket{1}) \to 
\frac{1}{\sqrt{2}}(\twoket{0}{0} + \twoket{1}{1}) .
\end{equation}
This correlation between the copies is an essential property for the
detection scheme presented here to be useful --- otherwise one could
not combine the results of the different detector measurements to
better infer properties of the original. For example the universal
quantum copying machine (UQCM)\cite{BuzekH:96}, which reproduces an
arbitrary qubit with the best possible fidelity cannot give gains in
detector efficiency via the scheme presented above, even when no
random noise is added in the copying process (analogous to $\varepsilon = 1$).
This matter will be further
investigated in Sec.~\ref{PROP} where the properties of the copying
machine required for this scheme to work are investigated.

\section{The Performance of the Copier-Enhanced Scheme With Noiseless Copiers}
\label{PERFORMANCE}

Firstly, consider the optimum case (for the copier-enhanced detection
scheme) when $\mu = -1$. In this situation, the copier produces a
vacuum when it fails to work, and
any noise present will
come only from the possibility of dark counts by the detectors. The
effect of copier noise will be considered in the next section, but for now we
will ignore it,
to show the general features of this
setup with greater clarity.

The detection scheme outlined in Sec.\ \ref{THEMODEL} provides the
observer who has the detectors with $2^N$ measurement
results, each of which can either be a ``count'' (henceforth labeled
as $+$), or ``no count'' (labeled as $-$). There are obviously better
and worse ways for the observer to use these $2^{(2^N)}$ distinct
possible outcomes to distinguish between a photon or vacuum input. Let
us look at two of these.

\subsection{Performance comparison for correctly choosing the most
likely input state}
\label{estimator}

 An obvious and simple  way to utilize the measurement results is to use them to
decide whether it is more likely that a photon or that vacuum was
input. One assumes that the person using the whole setup knows the
parameters $\eta, \xi, \varepsilon, \mu$. In statistical terminology, we
find the maximum likelihood estimator $\hat{\theta}$ for the parameter
$\theta$ which describes the input state $\ket{\theta}$, and so takes
on either the value $0$, or $1$. 

We wish to compare how well this strategy works with the
copier-enhanced scheme and with the basic one-detector setup. To this
end, we will compare $Q$, the probability that this ``most likely''
guess for the input state
(i.e. that $\hat{\theta} = \theta$) is correct.
For simplicity and clarity,
 we will restrict the
analysis of this method to the usual photodetection case when dark counts are rare \mbox{($\xi \ll 1$)}.

Consider first the standard detector-only setup (\mbox{$N=0$}). The measurement outcome probabilities
$P_{j|i}$  
[where $P_{j|i}$ is the probability of getting measurement result \mbox{$j \in \{+,-\}$},
given that the incident state was the $i$th one ($i \in \{0, 1\}$)] are easily found using Eqs.\ (\ref{Pi}) and\ (\ref{DETpovm})
\begin{mathletters}\label{N0probs}
\begin{eqnarray}
P_{+|1} =\eta,&\qquad& P_{-|1} =1-\eta,\\
P_{+|0} =\eta\xi,&\qquad& P_{-|0} =1-\eta\xi.
\end{eqnarray}\end{mathletters}
Now, the estimator $\hat{\theta}(j)$, given a certain measurement
result $j$, can be easily calculated from these, since
$\hat{\theta}(j) = i$ iff $P_{i|j} \ge 1/2$. One finds, for example,
that if a count is detected, then the most likely input was a photon
[$\hat{\theta}(+) = 1$] only if $p > \xi/(\xi+1)$. Similarly, the other
``common sense'' conclusion that if no count is seen, then it is more
likely that there was no input photon   [$\hat{\theta}(-) = 0$] occurs
only if $p < (1-\eta\xi)/[2-\eta(1+\xi)]$. This is because when $p$,
the probability of photon input is almost certain, then even if you
don't see it, it becomes more likely that an incoming photon wasn't
detected than that none came in at all. Let us ignore such situations when
$\hat{\theta}(+) = \hat{\theta}(-)$, since then this method tells us
nothing about the input state. The situation $\hat{\theta}(+)=0,
\hat{\theta}(-)=1$ never occurs. We find that for useful parameters,
the probability of being correct is 
\begin{eqnarray}
Q(0) &=& P_{+|1}p  + P_{-|0}(1-p) \nonumber\\
&=&  1-p +\eta[p - \xi(1-p)].
\end{eqnarray}

Now we want to compare to this the probability of being correct if some
quantum copiers are used to help things along. 
Consider the setup with only one copier
($N = 1$). The measurement outcome probabilities (where $P_{jk|i}$ is
the probability that given the $i$th input state, the first detector
gives the result $j$, and the second detector gives the result $k$),
are found using Eqs.\ (\ref{Pi}) - (\ref{COPYtransf}), remembering that
$\mu=-1$.
\begin{mathletters}\label{N1probs}
\begin{eqnarray}
P_{++|1} &=& \eta^2[\varepsilon+(1-\varepsilon)\xi^2],\\
P_{--|1} &=& \varepsilon(1-\eta)^2+(1-\varepsilon)(1-\eta\xi)^2,\\
P_{+-|1} = P_{-+|1}&=&\varepsilon\eta(1-\eta)+ (1-\varepsilon)\eta\xi(1-\eta\xi),\\
P_{++|0}&=&\eta^2\xi^2,\\
P_{--|0}&=&(1-\eta\xi)^2,\\
P_{+-|0} = P_{-+|0}&=&\eta\xi(1-\eta\xi).
\end{eqnarray}
\end{mathletters}

In this case, we find that the estimation method used in this
subsection is useful when $\hat{\theta}(++)=1$ and
$\hat{\theta}(--)=0$. This occurs when 
\begin{mathletters}\label{nogood}
\begin{equation}
  p > \frac{\xi^2}{\varepsilon(1-\xi^2)+2\xi^2}
\end{equation}
and
\begin{equation}
p < \frac{(1-\eta\xi)^2}{2(1-\eta\xi)^2-\varepsilon\eta[2-\eta(1+\xi)](1-\xi)},
\end{equation}
\end{mathletters}
respectively. Given these restrictions, there are still two
possibilities: when the results $(+-)$ or $(-+)$ are obtained, either
a photon or a vacuum input are more likely. It turns out that when the
vacuum is more likely in this situation
\mbox{[$\hat{\theta}(+-)=\hat{\theta}(-+)=0$]}, the $N=1$ detection scheme
with the copier always gives a worse probability of success than just
using a single detector ($N=0$). 

However, in the other case, when any
count on either of the detectors is more likely to indicate that a
photon was input, the scheme with the copier is often better. The
probability of a correct guess is then 
\begin{equation}
Q(1) = 1-p-\eta\xi(1-2p)(2-\eta\xi) +\varepsilon\eta p[2-\eta-\xi(2-\eta\xi)].
\end{equation}
And so,
 the copier-enhanced scheme gives better results whenever 
$Q(1)>Q(0)$, i.e. when
\begin{equation} 
  \varepsilon > \frac{\xi(1-\eta\xi)(2p-1)-p(1-\xi)}{p(1-\xi)[\eta(1+\xi)-2]}.
\end{equation}
In particular, in the usual practical situation with few dark counts
( $\xi\ll1$ ), and when the probability of photon input is much
greater than the probability of a dark count ($\xi\ll p$), this simplifies to 
\begin{equation}\label{guesscond}
  \varepsilon \gtrsim \frac{1}{2-\eta}
\end{equation}

So, the copier has to be just above 50\% efficient if the quantum efficiency $\eta$
of the detectors is low,
and somewhat better when $\eta$ is larger.

\subsection{Performance comparison for information about the initial state}
\label{IM}
  It was seen in the previous section that if one intends to make a definite
judgment about whether a photon was incident on the (single) detector or not, then
for some parameter values, the measurement result is no help at all. This is because
for these parameter values, the most likely original state is always
the same one, irrespective of the measurement result happens to be.
 The parameters $\eta$, $\xi$, and $p$ for which this is the case when
$N=1$ are those that
do not satisfy the relations\ (\ref{nogood}).

Nevertheless, in such a situation the fact that a count on a photodetector is still more likely (since $\xi<1$)
when the input is a photon then when the input is vacuum, means that this measurement
will always give at least \emph{some} information about what the input was.
(Of course if dark counts are very common, it will give only a minute amount).
It follows then, that the method of interpreting the results described in the
previous section\ \ref{estimator} (choosing the most likely possibility)  must be wasting some information about the input state.

Let's look instead at the total amount of information about the input state that is contained in
the measurement results. This is the (Shannon) mutual information $I_m$ per input state
 between some observer $A$ who knows with certainty what the
 original states are (perhaps because they were prepared by that
observer), and another observer $B$ who has access to the measurement
results of the detection scheme.
This can be readily evaluated from the expression \cite{Shannon:48a,Shannon:48b,Hall:97}
\begin{equation}\label{Imexpr}
  I_m = \sum_{i,j} P_{j|i} P_i \log_2 \frac{P_{j|i}}{P_j},
\end{equation}
where $i$ ranges over the number of possible input states, and $j$
over the number of possible detection results.
$P_i$ are the \emph{a priori} probabilities that the $i$th input state
entered the detection scheme, $P_{j|i}$ is the probability that
the $j$th the detection result was obtained given that the $i$th state
was input, and $P_j$ is the marginal probability that the $j$th
detection result was obtained overall.
In our case, the probabilities are given by\ (\ref{N0probs}) and (\ref{N1probs}),
and can be similarly readily evaluated for $N>1$. Actually, this calculation is
usually quite convenient, and avoids some of the complex formulae
encountered with the previous method of Sec.\ \ref{estimator}.
Also, unlike the latter, it is applicable for all parameter values.

 This mutual information has very concrete meaning even though in general, $B$ can never be
actually certain what any particular input state was. It is known that
by using appropriate block-coding and  error-correction schemes, $A$ can transmit to $B$
an amount of \emph{certain} information that can come arbitrarily close to
the upper limit $I_m$ imposed by the detection probabilities. In other words,
$I_m$ is the maximum amount of
information that $A$ and $B$ can share using a given detection scheme,
if they are cunning enough.

  It follows then, that the detection scheme which gives a greater information
content about the initial state $I_m$, will potentially be the more useful one.
The authors have shown that
 the Wootters-Zurek ($\varepsilon=1$) copier is the optimal
quantum broadcaster of information when the information is decoded one-symbol at a
time \cite{Us}.

From expression\ (\ref{Imexpr}) it can be seen that $I_m$ depends on the \emph{a priori} input
probabilities (the parameter $p$ in the cases considered here). This
leads one to surmise that (at least in general) various detection
schemes may do relatively better or worse depending on how frequently
the input is a photon. This is in fact found to be the case.
However, in what follows, we will concentrate mainly on the $p=1/2$ case of
equiprobable photons and vacuum, since
this is the situation that allows the maximum amount of information
to be encoded in the original message, and so is in some ways the most
basic case.

Lastly before we begin analyzing the new detection scheme, since $I_m$
becomes very small when most inputs are of the
same type (mostly photons, or mostly vacuum), it is convenient to
introduce an effective efficiency $\eta^e$ of the detection
scheme. If the new detection scheme gives mutual information content
$I_m(\varepsilon,\eta,\mu,\xi,N,p)$ per input state, then  $\eta^e(I_m(\varepsilon,\eta,\mu,\xi,N,p))$ is defined
as the efficiency of a noiseless detector  that would give the same mutual
information content if
it was used by itself in the basic scheme with no copiers. i.e.
\begin{equation}\label{nedef}
I_m(\cdot,\eta^e,\cdot,0,0,p) = I_m(\varepsilon,\eta,\mu,\xi,N,p).
\end{equation}
  $\eta^e$ is a one-to-one, monotonically increasing function of
$I_m$, and so if (and only if) some detection scheme increases $\eta^e$, it also
increases the mutual information. Thus, $\eta^e$ and $I_m$ are
equivalent for ranking detection schemes in terms of effectiveness.
$\eta^e$ also has the advantage that for some cases of the new
copier-enhanced detection scheme it is
independent of the photon input probability $p$. 
(Notably the basic
noiseless ( but possibly inefficient) case when $\mu=-1$, and $\xi=0$)

Now it is time to ask the question: for what parameter values does the
copier-enhanced detection scheme provide more information about the initial
states than using a single detector?

Consider firstly the simplest case of interest, where there are no
spurious (``dark'') counts in the photodetectors ($\xi=0$), and one has a copier of efficiency
$\varepsilon$, that produces vacuum upon failure ($\mu = -1$). This will give
some idea about the relationship between the detector and copier
efficiencies required, leaving the effects of noise for later consideration in
Sec.\ \ref{NOISE}.

As  mentioned previously, in
this situation the effective efficiency is independent of $p$, and
with one layer of copiers ($N=1$), it is found to be given by the simple
expression
\begin{equation}\label{effeff}
\eta^e_{(1)} = \varepsilon\left[1-(1-\eta)^2\right].
\end{equation}
Since this is independent of $p$,  introducing a second lot of copiers,
is equivalent to  replacing $\eta$ in the above expression by
$\eta^e_{(1)}$ i.e. $\eta^e_{(n+1)} =
\varepsilon\left[1-(1-\eta^e_{(n)})^2\right]$. In fact, in the limit of
never-ending amounts of copiers, the effective efficiency approaches
\begin{equation}\label{ninf}
\lim_{N\to\infty} \eta^e_{{}_{(N)}}  = 2 -\frac{1}{\varepsilon}.
\end{equation}
One finds that effective efficiency
 is improved (over \mbox{$\eta^e=\eta$}) by the copier scheme whenever
\begin{equation}\label{simplecond}
\varepsilon > \frac{1}{2-\eta}.
\end{equation}
This is the same as the condition\ (\ref{guesscond}) that is needed to improve the probability of
 making a correct guess with the method of Sec\ \ref{estimator}.

\begin{figure}
\center{\epsfig{figure=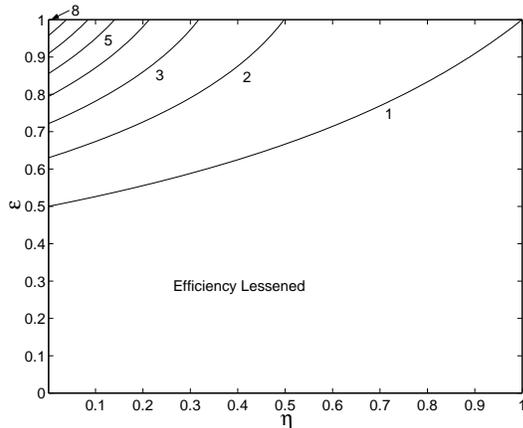,width=70mm}}
\caption{Relative efficiency gain $\eta^e/\eta$
contours  for the three-level ($N=3$) copier detection scheme over the
basic detector ($N=0$), as a function of detector efficiency $\eta$ and copier
efficiency $\varepsilon$, where both detectors and copiers are noiseless
($\xi=0,\mu=-1$). Valid for any photon input probabilities $p$.}
\label{nen3FIG}
\end{figure}

Since no random noise is introduced by either copier or detector,
improvement is achieved whenever more copiers are added, to arbitrary
order $N$. The relative improvement in effective efficiency
($\eta^e_{{}_{(3)}}/\eta$) when three layers of copiers are used ($N=3$)
is shown in Fig.~\ref{nen3FIG}.  A few things of interest to note in this
figure: \begin{enumerate}
\item The copier efficiency required is always above $\eta$ and above
$1/2$.
\item A gain in efficiency can be achieved even with quite poor copiers
--- for relatively small detector efficiencies $\eta$ (which occur for
photodetection in practice), the copier efficiency required is only
slightly above half.
\item For very good detectors, to get improvement, the copier efficiency $\varepsilon$ has to be
slightly greater than the detector efficiency $\eta$.
\item For low efficiencies, the relative gain in efficiency can be
very high, and can reach approximately $2^N$ for very poor detectors
and very good copiers.
\end{enumerate}

\begin{figure}
\center{\epsfig{figure=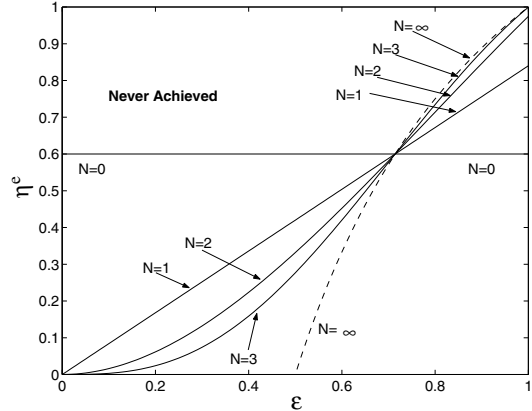,width=70mm}}
\caption{ Equivalent efficiency $\eta^e$, as a
function of copier efficiency $\varepsilon$ and number of levels of copiers
$N$, when detector efficiency is $\eta=0.6$, and both detectors and
copiers are noiseless ($\xi=0$, $\mu=-1$).
  Results for $N=0$ to $N=3$ are shown as solid lines, and the limit
of what can be achieved is shown as a dashed line. Regions beyond the
$N=0$ and $N\to\infty$ cases are not achievable with noiseless copiers.
Valid for all photon input probabilities $p$.}
\label{eta0.6FIG}
\end{figure}

To examine how much improvement can be achieved in more detail,
consider when the efficiency of the detectors is $\eta=0.6$. This is a
typical efficiency for a pretty good single-photon detector at present.
This
is shown by the solid lines in Fig.~\ref{eta0.6FIG}. Note how quite
large efficiency gains are achievable even when the copier efficiency
is slightly over the threshold useful value of $5/7 \approx 0.714$ (from
Eq.\ (\ref{simplecond})), and
how adding more copiers easily introduces more gains at first, but
after three levels of copiers, adding more becomes a lot of effort for not
much gain.

\section{The effect of Random Noise on Detection Scheme Usefulness}
\label{NOISE}

Following on from the analysis in Sec.\ \ref{IM}, let us now
 introduce various types of noise into the detection scheme.
 Unfortunately nice analytical results like\ (\ref{effeff}) - (\ref{simplecond})
 disappear, so what follows is based on the
results of numerical calculations. Additionally, the results now also depend on
the photon input probabilities $p$.  

\begin{figure}
\center{\epsfig{figure=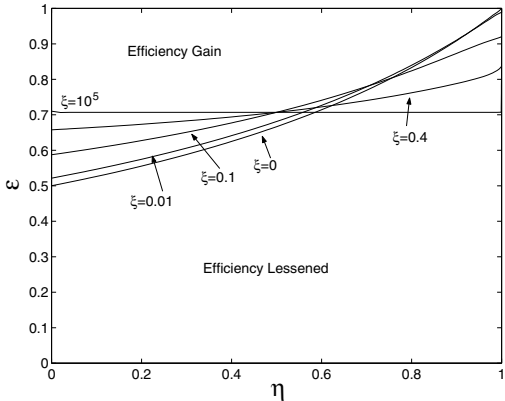,width=70mm}}
\caption{Regions of efficiency gain for the single-copier
($N=1$)
detection scheme, as a function of $\varepsilon$, the copier,  and
$\eta$,  the detector efficiencies, for varying frequency of dark counts parameterized
by $\xi$. In all cases, the copier produces vacuum when it fails
($\mu=-1$) and the input is equiprobable to be a photon or vacuum
($p=1/2$).}
\label{xiFIG}
\end{figure}

Firstly consider the effect of dark
counts ($\xi \neq 0$), while still keeping the copier noiseless ($\mu
= -1$). The regions of efficiency gain and loss with one copier are
shown in Fig.~\ref{xiFIG}. 
In real detectors, dark counts always occur, but are usually kept quite rare, so
realistic values of $\xi$ are of the order of $\xi \leqslant 0.01$.
 Thus (as
can be seen from Fig.~\ref{xiFIG}) for likely parameters, dark counts
do not reduce  the effectiveness of the copier detection scheme by
much at all. 

\begin{figure}
\center{\epsfig{figure=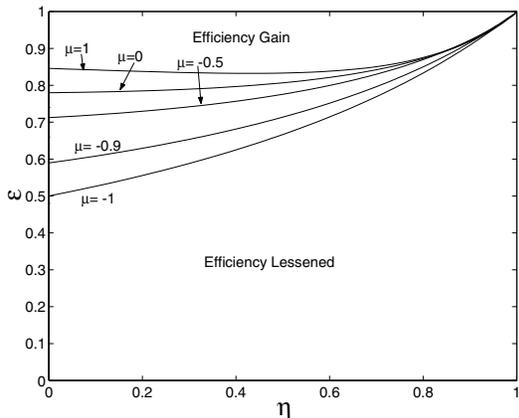,width=70mm}}
\caption{Regions of efficiency gain for the single-copier
($N=1$) detection scheme, as a function of $\varepsilon$, the copier, and
$\eta$, the detector efficiencies, for varying outputs when the copying fails ---
i.e. the variation in $\mu$.
 In all cases, the probability of dark counts in the detectors is
taken to be zero ($\xi=0$), and 
 the input is equiprobable to be a photon or vacuum ($p=1/2$).
  The case $\mu=-1$ corresponds to vacuum output upon failure of
copying, $\mu=0$ random output, $\mu=1$ photon output.}
\label{muFIG}
\end{figure}

Next, consider noise in the copier. In our scheme, noise is linearly introduced into the
copying process by varying the parameter $\mu$ away from $|\mu|=1$.
The amount of noise increases as $\mu$ approaches zero, until only  pure noise
 occurs  upon
failure of the copying for $\mu=0$. The dependence on $\mu$ of the
values of $\varepsilon$ and $\eta$ needed for efficiency gain is shown in
Fig.~\ref{muFIG}. In the particular case shown, photons and vacuum
are equiprobable ($p=1/2$) and there are no dark counts ($\xi=0$).

Firstly, it is seen that in most cases\cite{phovac}, the optimum output for the copier to produce
upon failure is vacuum ($\mu=-1$), and the worst situation is when it
produces photons by default ($\mu=1$). Totally random  default output
($\mu=0$) requires the copier inefficiency to be reduced by roughly a
factor of two relative to what is permissible for vacuum default
output. Unfortunately, little is known to date about how much noise will
be inevitably introduced in a practical quantum copier, but it seems
reasonable that the default output can be made somewhat (perhaps
significantly) better than
random. If noise could be made 10\% probable (perhaps not an unreasonable figure) upon failure to copy,
then copiers with efficiency $\varepsilon$ of about $0.65$ would improve detection
for typical quantum efficiencies $\eta$ of about $0.3$ or $0.4$. Either way, it
is seen that even overwhelming noise upon failure to copy, still
allows fairly inefficient (say $\varepsilon\approx 0.8$) copiers to improve
the detection efficiency. This is perhaps somewhat unexpected.

\begin{figure}
\center{\epsfig{figure=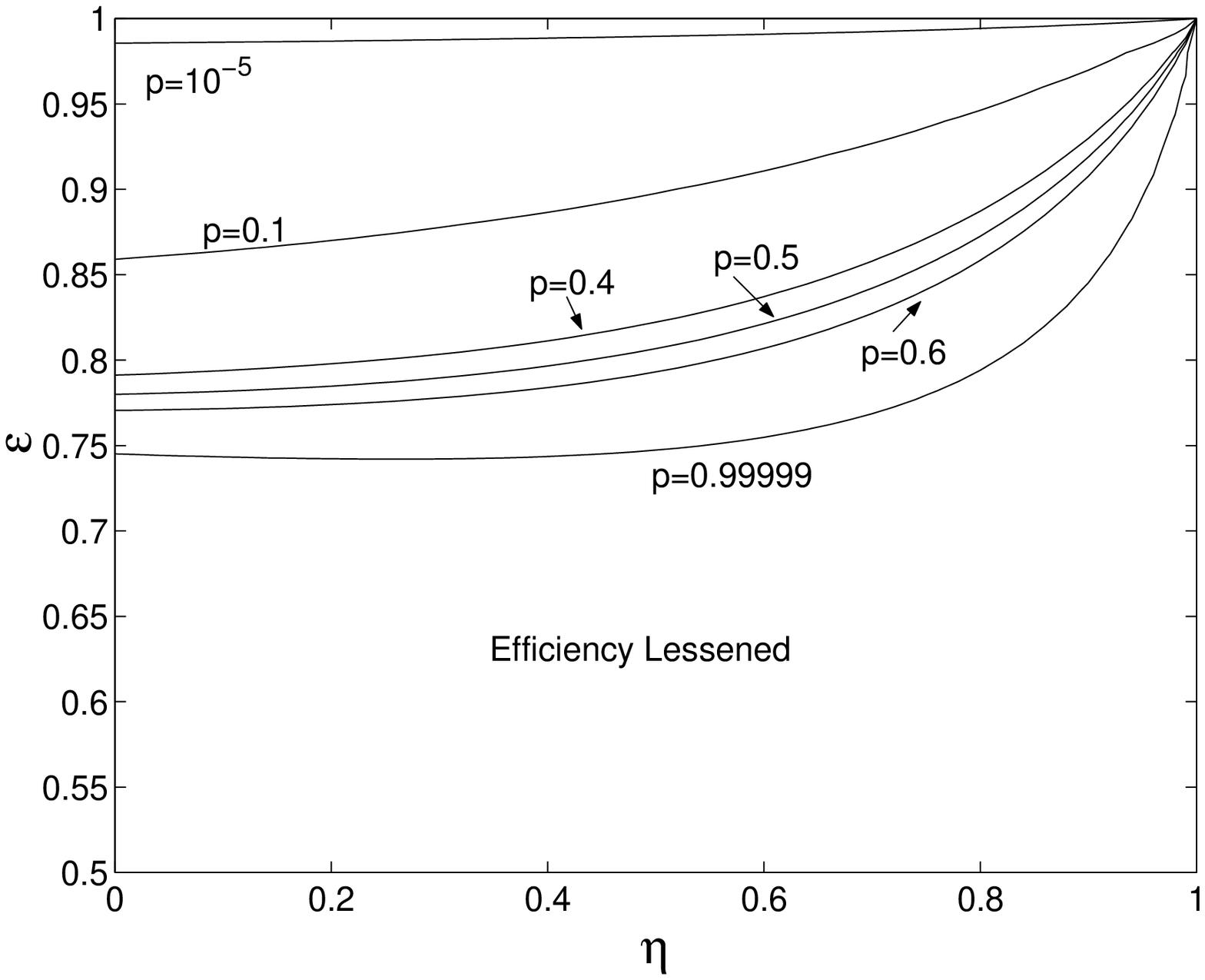,width=70mm}}
\caption{Regions of efficiency gain for the single-copier
($N=1$) detection scheme, as a function of $\varepsilon$, the copier, and
$\eta$, the detector
efficiencies, for varying \emph{a priori} photon input
probabilities $p$. 
 In all cases, the probability of dark counts in the detectors is
taken to be zero ($\xi=0$), and the default output upon copy failure
is a totally random state ($\mu=0$). Note that the scale in $\varepsilon$
differs from that in Figs.~\ref{nen3FIG},~\ref{xiFIG}, and~\ref{muFIG}}
\label{pFIG}
\end{figure}

Since the effective efficiency $\eta^e$ only varies with photon
frequency $p$ when noise is present, the next question which arises when
considering noisy schemes, is what effect does $p$ have on the
performance of the new detection scheme?
Fig.~\ref{pFIG} shows regions of efficiency increase in terms of
$\varepsilon$ and $\eta$ for a single
copier scheme, when
it is used on sets of input states containing different proportions of
photons $p$. The copier in this case produces the maximum amount of noise
upon failure (i.e. $\mu=0$).
Features seen include
\begin{enumerate}
\item Efficiency is easiest to increase when $p$ is close to one,
i.e. there are photons coming in most of the time.
\item When photons are rare ($p$ small), the copiers have to be very
efficient to be useful, since one wants to register almost all of those
that do come along.
\item When photons and vacuum are of a similar frequency, the necessary
copier efficiency changes slowly. (see how the $p=0.4,0.5,0.6$ curves
are close together).
\end{enumerate} 
 The
behaviour exhibited is fairly typical, although $\mu=0$ appears to be the
worst case scenario, as it is the most noisy.
In less noisy
situations, the required copier efficiency $\varepsilon$ increases
more slowly with decreasing $p$.
\section{Required Copier Properties}
\label{PROP}
As mentioned briefly at the end of Sec.~\ref{THEMODEL}\ (\ref{superpos}), the fact
that the quantum copier produces entangled states when the input is a
superposition is important for the scheme outlined above to work. Let
us consider what properties a quantum copier must have to be useful in
this scheme.

   The scenario where it is easiest to enhance the detection of
information is where the detectors are very weak ($\eta$ very small)
and there are no dark counts ($\xi = 0$). So, if a copier is of no use
in this situation, it will not be useful for any detector parameters
whatsoever. This will let us specify the broadest range of copier
parameters for which they may be useful in improving detection
efficiencies.  

In any detection situation, we can choose the basis in which to
specify the transformation to be the one in which the detectors measure
populations only. Since this simplifies the mathematics, let us do
so in what follows. 
We impose one condition on the
copier to make the analysis clearer: the states of the copies
considered separately (that is, the reduced density matrices of the
copies)  are identical. This is the usual situation, where both copies
are the same.
This allows us to write the copying
transformation of the two possible input states (including any noise introduced
by experimental factors) as
\begin{mathletters}\begin{eqnarray}
\ket{1}\bra{1} \to&
a_1\twoket{1}{1}\twobra{1}{1}+a_2\twoket{0}{0}\twobra{0}{0}\nonumber\\
&+\frac{1}{2}(1-a_1-a_2)\op{M}
 + \op{C}_a, \\
\ket{0}\bra{0} \to&
b_1\twoket{1}{1}\twobra{1}{1}+b_2\twoket{0}{0}\twobra{0}{0}\nonumber\\
&+\frac{1}{2}(1-b_1-b_2)\op{M}
+ \op{C}_b, 
\end{eqnarray}
where 
\begin{equation}
\op{M} =
\twoket{0}{1}\twobra{0}{1}+\twoket{1}{0}\twobra{1}{0},
\end{equation}
\end{mathletters}
and where \mbox{$0 \le a_1+a_2 \le 1$}, \mbox{$0 \le b-1+b_2 \le 1$}, and $\op{C}_a$,
$\op{C}_b$ consist only of coherences, so do not contribute to the
measurement probabilities, since we have chosen the basis so that what
the detectors measure become the populations.

The information about the original states transmitted to the observer with the detectors can be easily calculated
using the relations\ (\ref{Pi}),\ (\ref{DETpovm}),\ (\ref{Imexpr}),
noting that when a copier is present, the POVM which describes the
combined measurements at both detectors
simply consists of all tensor products of the one-detector POVM
\begin{equation}
  \{\op{A}_{nm} = \op{A}_n\otimes\op{A}_m\ :\ n,m \in \{+,-\}\}. 
\end{equation} 
The information with weak detectors, input photon probability $p$, and no copier is
\begin{equation}
  I_o(\eta,p) = -\eta p\log_2 p.
\end{equation}
And with copier:
\begin{eqnarray}\label{ABIm}
 I_m(\eta,&p&,A,B)\nonumber\\ &=&\eta pA\log_2 A + \eta(1-p)B\log_2 B\nonumber\\ &&-
 \eta[pA+(1-p)B]\log_2 [pA+(1-p)B],
\end{eqnarray}
that depends only on two parameters of the copier
\begin{equation}
A = 1+a_1-a_2 \quad;\quad B = 1+b_1-b_2.
\end{equation}

\begin{figure}
\center{\epsfig{figure=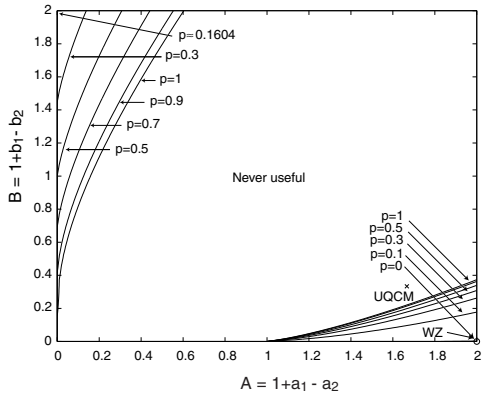,width=70mm}}
\caption{Quantum copier properties which allow improvements in
information transfer when using detectors of very low efficiency
$\eta$ having no dark counts ($\xi=0$). 
 $A$ and $B$ are parameters of
the copiers, and the lines show boundaries of the regions in
$(A,B)$ space within which copiers give improvements. Various lines
correspond to various \emph{a priori} input photon probabilities $p$ indicated on the
plot. Improvements occur in the regions away from the diagonal
$A=B$ relative to the boundary lines for a given $p$. For higher efficiencies
$\eta$, smaller regions of $A-B$ parameter space are useful.
The UQCM is indicated by the cross at $(A,B) =(5/3,1/3)$, 
and the Wootters-Zurek copier by the
circle at $(2,0)$.}
\label{copFIG}
\end{figure}

Fig.~\ref{copFIG} shows the values of parameters $A$ and $B$ over
which copiers are useful for detection enhancement, for various
$p$. Some points to note about this figure:
\begin{enumerate}
\item The diagonal $A=B$ corresponds to (via\ (\ref{ABIm}))  the worst-case situation where no
information about the input states is recoverable from the detectors
($I_m=0$).
\item When $A>1$, photon inputs create photon outputs more often than
vacuum, while if $B<1$, vacuum inputs create vacuum outputs more often
than photons. Thus, the region $A>1$,$B<1$ corresponds to imperfect
cloning transformations, while the region $A<1$,$B>1$ corresponds to 
imperfect 'swapping' transformations which most often transform photons into
vacuum, and vacuum into two photons.
\item Relabeling $\ket{1} \to \ket{0}$, and $\ket{0} \to \ket{1}$ in
the copying transformation does not keep the recovered information
$I_m$ invariant because the detectors
do not react the same way to photons and vacuum. This is why
Fig.~\ref{copFIG} is not symmetric about $A=B$.
\item The noisy copying transformation\ (\ref{COPYtransf}) used in previous sections of
this article can be made to correspond to any values of $A$ and $B$
where $A>B$ by appropriate choices of $\mu$ and $\varepsilon$. In fact,
\begin{mathletters}
\begin{eqnarray}
A &=& 1+\mu +\varepsilon(1-\mu), \\
B &=& 1+\mu -\varepsilon(1+\mu),
\end{eqnarray}
\end{mathletters}
and families of such transformations with a set efficiency $\varepsilon$ are
parallel to the dividing line $A=B$.
\item
  Greater ranges of copiers become useful as $p$ (the \emph{a priori}
  input photon frequency) becomes larger. For very low photon frequencies,
  only the close vicinity of $(A,B) = (2,0)$ gives improvements.
\item
  The Wootters-Zurek copying machine (or entangler) lies at this point
  $(A,B)=(2,0)$, and is the only copying transformation which gives
  improvement for arbitrary photon frequency $p$.
\item
  The well known Universal Quantum Copying Machine (UQCM)\cite{BuzekH:96}, which reproduces an
  arbitrary qubit with the best fidelity lies at $(A,B)
  =(5/3,1/3)$, outside the region of detection
  improvement for any $p$, and hence is never useful for the type of
  detection enhancement scheme discussed here.
\end{enumerate}

Thus one can see that quantum copying transformations used in such a
detection improvement scheme as outlined here must be similar in their
properties to the Wootters-Zurek copier (the controlled-NOT gate), and
the degree of similarity required depends on the input photon frequency.

\section{Conclusions}
\label{CONC}
We have provided an example of how spreading information about quantum
states onto a larger number of subsystems, actually increases the
amount of information about the original state that is available to an
observer. The key reason why this occurs is that in realistic situations, observers are
always restricted in how close to the ideal their measurements can
be.  Then, quantum copying the original state may allow the observer to make better use
of the detection apparatus at their disposal. 

In particular,
more efficiency of detection can be gained by employing entangling quantum
copiers such as a controlled-NOT gate. In fact if the efficiency of the detectors
is far from 100\% (such as in single-photon detection) the copier does not
have to be very efficient itself, and significant gains in detection
can still be made. 

From Fig.~\ref{nen3FIG}, and others, it can be
seen that to be useful, the quantum copiers must be successful with an
efficiency $\varepsilon$ over 50\% and somewhat greater than the detector
efficiency $\eta$. It is not generally clear how feasible this is for
 various physical systems, or measurement schemes that one might
wish to employ. With current technology it is often still
easier to make measurements on a system, rather than entangling it
with other known systems, however this varies from measurement to measurement
and from system to system. The physical processes involved in
measurement and quantum copying are often quite different: the former
requires creating a correlation between a quantum system and a
macroscopic pointer, whereas the latter involves creating quantum
entanglement between two similar microscopic states. Efficient
detection depends on correlating the system with its environment in a
strong, yet controlled way, whereas quantum copying depends on
isolating the system from its environment. One thus supposes that the
usefulness of a scheme such as the one outlined here will depend on
the system and measurements in question, due to the relative ease of
implementing detection and controlled quantum evolution in those systems.

 The copier parameters required for usefulness of the proposed scheme when random
noise is present are found to depend somewhat on the relative frequency of the
various states to be distinguished.  In any case, the copying
transformation must be similar to a controlled-NOT gate, the exact
degree of similarity depending on the relative frequency of the input states.
The effectiveness of the scheme is, however,
quite robust to random noise in the detection and copying.
We note that although a
detailed analysis was carried out for the case of single-photon detection,
the basic scheme immediately generalizes to the case of distinguishing
between any two mutually orthogonal states with inefficient detectors, and
can be readily generalized to a larger set of input states, and different
detectors.

The analysis that is carried out 
in terms of mutual information between the sequence of input states,
and an observer using the detection scheme, 
is seen to be a  simple to use, and powerful method of evaluating
detection schemes.

\begin{acknowledgements}
  We are grateful to R., P., and M. Horodecki for an illuminating discussion, 
  and we appreciate the helpful remarks from an anonymous referee
  regarding Sec.~\ref{estimator}.
\end{acknowledgements}
 

\end{multicols}


\begin{references}

\bibitem[*]{PDemail} Email address: deuar@physics.uq.edu.au

\bibitem{summ}
P. Deuar and W.~J. Munro, Phys. Rev. A {\bf 61}, 010306(R) (2000).

\bibitem{WZ:82}
W.~K. Wootters and W.~H. Zurek, Nature (London) {\bf 299},  802  (1982).

\bibitem{Barnumetal:96}
H. Barnum, C.~M. Caves, C.~A. Fuchs, R. Jozsa, and B. Schumacher, Phys. Rev.
  Lett. {\bf 76},  2818  (1996).

\bibitem{Kraus:83}
K. Kraus, {\em States, Effects, and Operations: Fundamental Notions of Quantum
  Theory} (Springer, Berlin, 1983).

\bibitem{CavesD:94}
C.~M. Caves and P.~D. Drummond, Rev. Mod. Phys. {\bf 66},  481  (1994).

\bibitem{BuzekH:96}
V. Bu\v{z}ek and M. Hillery, Phys. Rev. A {\bf 54},  1844  (1996).

\bibitem{Shannon:48a}
C.~E. Shannon, Bell Syst. Tech. J. {\bf 27},  379  (1948).

\bibitem{Shannon:48b}
C.~E. Shannon, Bell Syst. Tech. J. {\bf 27},  623  (1948).

\bibitem{Hall:97}
M.~J.~W. Hall, Phys. Rev. A {\bf 55},  100  (1997).

\bibitem{Us}
P. Deuar and W.~J. Munro, Phys. Rev. A (to be published).

\bibitem{phovac}
For high detector efficiencies $\eta$, when dark counts
occur, photon default output upon failure to copy ($\mu=1$) may be
more desirable than vacuum.
\end{references}
\end{document}